\documentclass[sort&compress]{elsarticle}
\usepackage{amsmath}
\usepackage{amssymb}
\usepackage{graphicx}

\newcommand{\ket}[1]{\left|#1\right>}
\newcommand{\bra}[1]{\left<#1\right|}
\newcommand{\abs}[1]{\left|#1\right|}

\begin{document}

\begin{frontmatter}

\title{Cavity-assisted energy relaxation for quantum many-body simulations}

\author[kias]{Jaeyoon Cho\corref{cor}}
\ead{choooir@gmail.com}
\author[ucl]{Sougato Bose}
\author[imperial,kias]{M. S. Kim}

\cortext[cor]{Corresponding author. Tel.: +82 2 958 3850}

\address[kias]{School of Computational Sciences, Korea Institute for Advanced Study, Seoul 130-722, Korea}
\address[ucl]{Department of Physics and Astronomy, University College London, Gower Street, London WC1E 6BT, UK}
\address[imperial]{QOLS, Blackett Laboratory, Imperial College London, London SW7 2BW, UK}

\begin{abstract} 
We propose an energy relaxation mechanism whereby strongly correlated spin systems decay into their ground states. The relaxation is driven by cavity quantum electrodynamics interaction and the decay of cavity photons. It is shown that by applying broadband driving fields, strongly correlated systems can be cooled regardless of the specific details of their energy level profiles. The scheme would also have significant implications in other contexts, such as adiabatic quantum computation and steady-state entanglement in dissipative systems. 
\end{abstract}


\end{frontmatter}

\section{Introduction}

Advances in laser cooling and trapping technologies
have opened up intriguing possibilities of using atomic systems to simulate
quantum many-body phenomena in strongly correlated
regimes~\cite{jaksch:1998a,greiner:2002a}. Concerning such quantum
simulation, diverse studies have been carried out both experimentally and
theoretically, addressing various fundamental many-body
problems~\cite{lewenstein:2007a}. From the theoretical perspective, the
properties of the ground states are among the main concerns regarding
strongly correlated systems. On the other hand, studies on the experimental
realizations have rather placed the main emphasis on how to mimic particular
many-body model Hamiltonians, while it is often overlooked that being able to
realize a particular Hamiltonian does not mean that its
energy states can be prepared.

The most natural way to bridge such a logical gap would be to introduce
energy relaxation in simulated strongly correlated systems. First of all, it is
important to understand that in strongly correlated regimes no conventional
cooling scheme provides a proper relaxation
channel~\cite{anderson:1995a,bradley:1995a,davis:1995a,metcalf:1999a,horak:1997a,maunz:2004a,larson:1986a,daley:2004a,griessner:2004a}. The main reason is because they
were devised to cool weakly correlated systems, keeping it in mind that
cooling individual atoms results in cooling the whole system. This does not
hold in strongly correlated regimes. Moreover, while important classes of
many-body systems emerge when the atomic internal degrees of freedom (DOFs),
i.e., spins, are strongly correlated, cooling of such systems is yet to be
considered. In weakly correlated regimes, optical pumping could be regarded
as the spin counterpart of the conventional cooling of atomic motions.
However, the problem again becomes elusive when the spins are strongly
correlated.

Up until now, the adiabatic method is the conventional way to prepare the ground state of a given strongly correlated system~\cite{jaksch:1998a, greiner:2002a}.
The underlying idea is to prepare a known ground state in a weakly correlated regime and transfer it to that in a desired strongly correlated regime by adiabatically changing the system parameter. 
This method is feasible when the ground-state
energy gap is finite throughout the whole process. However, this condition is
difficult to achieve because such a transition (from a weakly correlated to a
strongly correlated regime) is usually a quantum phase transition, at the
critical point of which the energy gap vanishes in the thermodynamic
limit~\cite{sachdev:1999a}. 
As adiabatic dynamics is a
slow unitary evolution, the coherence time of the system also limits its
performance both during and after the preparation, whereas energy relaxation
would continuously stabilise the ground state competing with decoherence.

In this paper, we introduce a cavity-assisted energy relaxation mechanism that directly cools strongly correlated atomic systems. 
As a concrete example, we discuss cooling of strongly correlated spin systems~\cite{duan:2003a,garcia-ripoll:2004a,sim11,porras:2004a,friedenauer:2008a,kim10,bri12,isl13,hartmann:2007a,cho:2008a}.
Although some schemes have been proposed to prepare particular ground states~\cite{rabl:2003a,popp:2006a,diehl:2008a,verstraete:2009a}, there has hitherto been no genuine idea of energy relaxation applicable to general many-body Hamiltonians whose ground states are unknown.
Besides the application in quantum simulations, the importance of energy relaxation can also be found in many other contexts. 
For example, it has a potential to provide a more efficient route to
reach the ground state for adiabatic quantum computation~\cite{farhi:2001a}.
Also, recalling that strongly correlated spin systems have entangled ground
states, it would provide a general framework for establishing 
steady-state
entanglement, discussed so far from different perspectives~\cite{diehl:2008a, verstraete:2009a}. 
Our scheme is relevant to current interests in placing many-body atomic systems inside cavities~\cite{mekhov:2007a, eckert:2008a,jiang:2008a,brennecke:2007a,colombe:2007a,wol12,bau10,rit13}.

\section{Physical model}

We consider strongly correlated spin systems as can be simulated in various atomic systems, such as optical lattices~\cite{duan:2003a, garcia-ripoll:2004a,sim11}, 
ion traps~\cite{porras:2004a, friedenauer:2008a,kim10,bri12,isl13}, and 
coupled cavities~\cite{hartmann:2007a, cho:2008a}. 
The common feature of those quantum
simulations is that atomic hyperfine ground levels represent spins, while
some external DOF (atomic momentum, cavity modes, etc.) mediates
their coupling. In so doing, the external DOF is effectively decoupled from
the {dynamics}. Our concern is how to manipulate 
{the} spin DOF to lower its
respective energy.

\begin{figure}
	\includegraphics[width=0.25\columnwidth]{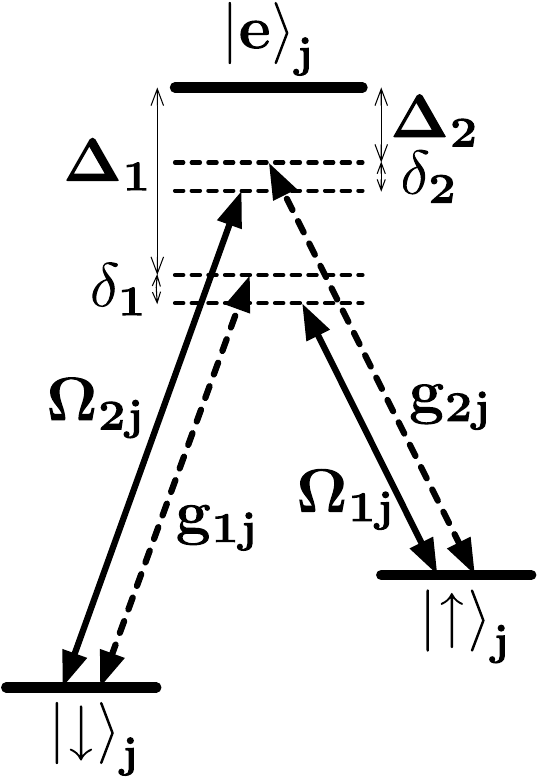}
	\caption{Atomic levels and transitions. Two transitions $\ket\downarrow_j\leftrightarrow\ket{e}_j$ and $\ket\uparrow_j\leftrightarrow\ket{e}_j$ are coupled, respectively, to the two orthogonally polarized cavity modes. Two Raman transitions represented by the subscripts 1 and 2 are embedded.}
	\label{fig:level}
\end{figure}

For {simplicity}, we present our idea with spin-$\frac{1}{2}$ systems. Two
ground levels of an atom represent spin-down and up states, as shown in
Fig.~\ref{fig:level}. To begin with, let us write the Hamiltonian of the
quantum simulator as
$H_0=\sum_{\mu=0}^{2^N-1}E_\mu\ket{\Psi_\mu}\bra{\Psi_\mu}$, where $E_\mu$ is
the $\mu$th eigenenergy with $E_\mu\le{E_\nu}$ for $\mu<\nu$,
$\ket{\Psi_\mu}$ is the $\mu$th eigenstate, and $N$ is the number of spins.
We assume that the ground state $\ket{\Psi_0}$ is nondegenerate. Our aim is
to figure out a process that allows cooling
($\ket{\Psi_\mu}\rightarrow\ket{\Psi_\nu}$ for $\mu>\nu$) but forbids heating
($\mu<\nu$). 

Note that in the aforementioned simulations,
spontaneous emission is not feasible as a relaxation channel for the following 
reasons: (i) As the spontaneous emission rate is generally larger than the
characteristic energy scale of the simulator, the energy levels can not be
resolved due to the line broadening. (ii) The spontaneous emission destroys
the atomic coherence, hence the coherences within the many-body states. (iii) As
atomic excited levels are out of the spin subspace in consideration, the
energy of the simulator is not clearly defined when the atom is excited. 
(iv) Given complicated energy profiles of strongly correlated systems, it is impractical (or even impossible for many atoms) to discriminate between cooling and heating transitions only by their respective frequencies. One requires a mechanism to obstruct heating regardless of the frequencies of driving fields.

For the above reasons, we need to introduce some ancillary DOF so that energy is
first transferred coherently to it and then relaxed from it. 
Let us denote by
$\ket{\Psi_\mu,n}$ the state of the system along with the ancillary mode in
the $n$th excited state. Our strategy is to find such a transition as
$\ket{\Psi_\mu,n}\leftrightarrow\ket{\Psi_\nu,n+1}
\rightarrow\ket{\Psi_\nu,n}$.
Provided that the former part of the transition preserves the energy of the whole system, a transition with $\mu>\nu$ is favored as the energy of $\ket{n+1}$ is larger than that of $\ket{n}$. 
Note that the irreversible
transition in the internal DOF is attained by destroying the coherence of the
external DOF rather than by destroying that of the internal DOF. 
It is worth comparing this with the sideband cooling {that
works in an opposite way}. Denoting by $\ket{g}$ and $\ket{e}$ the atomic
ground and the excited states, respectively, the sideband cooling can be
represented as $\ket{n,g}\leftrightarrow\ket{n-1,e}\rightarrow\ket{n-1,g}$.

It can be seen that the cavity field is best suited for the above purpose.
Among various possibilities, we consider a situation where one cavity is
coupled to the quantum simulator, as depicted in Fig.~\ref{fig:level}. Two
transitions $\ket\downarrow_j\leftrightarrow\ket{e}_j$ and
$\ket\uparrow_j\leftrightarrow\ket{e}_j$ of the $j$th atom are coupled,
respectively, to the two orthogonally polarized cavity modes, whose
annihilation operators are denoted by $a_1$ and $a_2$, with coupling rates
$g_{1j}$ and $g_{2j}$ and detunings $\Delta_1$ and $\Delta_2$. The classical
field with Rabi frequency $\Omega_{xj}$ ($x=1,2$) is also applied with
detuning $\Delta_x+\delta_x$. We take a large detuning regime where
$\Delta_x,\Delta_1-\Delta_2{\gg}g_{xj},\Omega_{xj},\delta_x$. In this regime,
the atomic excitation is suppressed and the conventional adiabatic
elimination method can be used to yield the effective
Hamiltonian~\cite{james:2007a}
\begin{equation} 
	\begin{split} 
		H_1&=H_d+H_t,\\
		H_d&=-\epsilon_{z}-\epsilon_1a_1^{\dagger}a_1
			-\epsilon_2a_2^{\dagger}a_2,\\
		H_t&=-(a_1^{\dagger}\Gamma_{-}e^{i\delta_1t}
			+a_2^{\dagger}\Gamma_{+}
			e^{i\delta_2t}+\text{H.c.}), 
	\end{split} 
	\label{eq:hamil1} 
\end{equation}
where
$\epsilon_{z}=\sum_{j}\left(\frac{\left|\Omega_{2j}\right|^{2}}{\Delta_{2}}-\frac{\left|\Omega_{1j}\right|^{2}}{\Delta_{1}}\right)s_{j}^{z}$,
$\epsilon_{1}=\sum_{j}\frac{\left|g_{1j}\right|^{2}}{\Delta_{1}}s_{j}^{-}s_{j}^{+}$,
$\epsilon_{2}=\sum_{j}\frac{\left|g_{2j}\right|^{2}}{\Delta_{2}}s_{j}^{+}s_{j}^{-}$,
$\Gamma_{-}=\sum_{j}\frac{g_{1j}\Omega_{1j}}{\Delta_{1}}s_{j}^{-}$, and
$\Gamma_{+}=\sum_{j}\frac{g_{2j}\Omega_{2j}}{\Delta_{2}}s_{j}^{+}$. Here, we
employed the spin {notation} with
$s^z=\frac{1}{2}(\ket{\uparrow}\bra{\uparrow}-\ket{\downarrow}\bra{\downarrow})$
and $s^+=(s^-)^\dagger=\ket{\uparrow}\bra{\downarrow}$.

\section{Working principle}

Let us take two of the energy eigenstates $\ket{\Psi_\mu}$ and $\ket{\Psi_\nu}$ of the spin system. 
In what follows, we adopt the convention that $\bra{\Psi_\mu}O\ket{\Psi_\nu}$ for any operator $O$ is denoted by $(O)_{\mu\nu}$. 
Our plan is to treat the Hamiltonian~\eqref{eq:hamil1} as a time-dependent perturbation. 
For this, we impose
$\abs{E_{\mu\nu}}\gg(\epsilon_z)_{\mu\mu},(\epsilon_z)_{\nu\nu},|(\epsilon_z)_{\mu\nu}|$,
and similarly for $\epsilon_1$, $\epsilon_2$, $\Gamma_\pm$, and
$\Gamma_\pm^\dagger$, where $E_{\mu\nu}=E_\mu-E_\nu$ (Condition 1).

The Hamiltonian~\eqref{eq:hamil1} comprises the diagonal term $H_d$ and the
off-diagonal term $H_t$. The former makes a small correction to the energy
levels, which we take into account later. The essential part is the latter
that induces transition between the energy levels. Let us denote by
$\ket{\Psi_\mu,n_1,n_2}$ the state with $n_1$ and $n_2$ photons in the $a_1$
and $a_2$ modes, respectively. If the cavity has initially no photon, the
transition occurs creating a cavity photon. The transition rate is thus
determined by the matrix elements $\bra{\Psi_\mu,1,0}H_I\ket{\Psi_\nu,0,0}
=(\Gamma_-)_{\mu\nu}e^{i(E_{\mu\nu}+\delta_1)t}$ and
$\bra{\Psi_\mu,0,1}H_I\ket{\Psi_\nu,0,0}
=(\Gamma_+)_{\mu\nu}e^{i(E_{\mu\nu}+\delta_2)t}$, where
$H_I=e^{iH_0t}{H_t}e^{-iH_0t}$ is the interaction Hamiltonian. This
expression indicates that the transition occurs dominantly when the resonance
condition $E_{\mu\nu}+\delta_{\{1,2\}}=0$ is satisfied. Furthermore, if
$\delta_{\{1,2\}}$ is chosen to be positive, the resonance condition is
satisfied only when $\mu<\nu$, which means that the transition occurs in
favor of decreasing the energy of the quantum simulator. $\delta_{\{1,2\}}$
being positive implies that the two-photon Raman transition takes place in
such a way that an atom absorbs a lower-energy photon from the classical
field and emits a higher-energy photon into the cavity mode. The net result
is thus that the amount of energy $\delta_{\{1,2\}}$ is transferred from the
quantum simulator to the cavity mode. The subsequent decay of the cavity
photon then completes one cycle of the energy relaxation.
The correction by $H_d$ in Hamiltonian~\eqref{eq:hamil1}, albeit small, can
be taken into account as follows. $H_d$ shifts the energy of state
$\ket{\Psi_\mu,n_1,n_2}$ as much as
$-(\epsilon_z+n_1\epsilon_1+n_2\epsilon_2)_{\mu\mu}$, which modifies the
resonance condition as
$E_{\mu\nu}-(\epsilon_z+\epsilon_{\{1,2\}})_{\mu\mu}+(\epsilon_z)_{\nu\nu}+\delta_{\{1,2\}}=0$.
This term also modifies the energy eigenstates themselves. The infidelity of
the state for $n_1=n_2=0$ due to this correction is of order
$\abs{(\epsilon_z)_{\mu\nu}/E_{\mu\nu}}^2$.

We impose the following additional conditions. (Condition 2) We require $\abs{E_{\mu\nu}}\gg\kappa$, where $\kappa$ is the cavity decay rate. Otherwise, the energy levels can not be resolved due to the line broadening, which results in heating of the quantum simulator. (Condition 3) For the lower energy state, $(\Gamma_\pm)_{\mu\mu}\ll\kappa$ should be satisfied, since otherwise the cavity photon can be generated without a transition, resulting in heating, e.g., through transition $\ket{\Psi_\mu,1,0}\rightarrow\ket{\Psi_\nu,0,0}$ that is allowed for $\mu<\nu$. 

\begin{figure}
	\includegraphics[width=0.45\columnwidth]{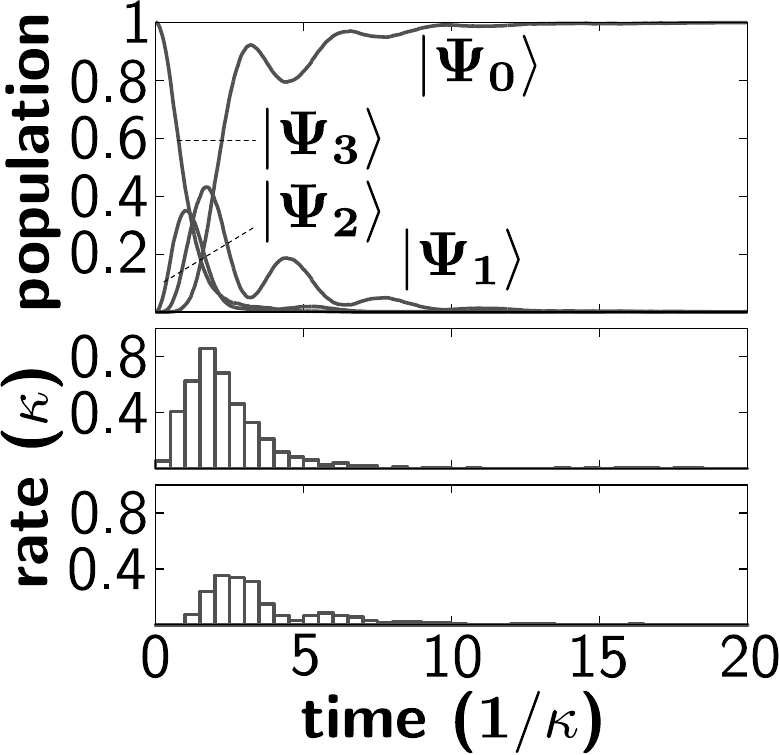}
	\caption{Numerical simulation for a two-spin system: time evolution of the population (top panel) and detection rates of the cavity photons (middle and bottom pannels).}
	\label{fig:toy}
\end{figure}

As an example, we analyze in Fig.~\ref{fig:toy} the energy relaxation process
of a two spin-$\frac12$ system with Hamiltonian
$H_0=B(s_1^z+s_2^z)+J\vec{s}_1\cdot\vec{s}_{2}$ and $B=2J>0$. In this
case $E_{\mu+1}-E_\mu=B$ for
every $\mu$. The parameters are chosen as
$\Delta_{\{1j,2j\}}/100=\Omega_{1j}/10=g_{\{1j,2j\}}=7\kappa$ and
$\delta_{\{1,2\}}=10\kappa=B$. $\Omega_{2j}$ is chosen to be $-\Omega_{1j}$
to optimize the process. $\epsilon_z$ is taken to be zero. The top panel of
Fig.~\ref{fig:toy} shows the population of each energy level with respect to
time and the middle and the bottom panels show, respectively, the detection
rates of the photons in mode $a_1$ and $a_2$ averaged over the corresponding
time bins by using the quantum trajectory method~\cite{carmichael:1993a}. The
population, initialized in the highest energy level to show the cascade-like
transition structure, eventually trapped in the ground level, during which
the detection ceases, indicating the completion of the relaxation.
An oscillatory behavior appears in Fig.~\ref{fig:toy} as the cavity photon 
can be reabsorbed into 
the atoms before leaking out of the cavity, e.g., as in the oscillation 
$\ket{\Psi_{\mu},0,0}\leftrightarrow\ket{\Psi_{\mu-1},1,0}$. Evidently, this
oscillation can persist only until the cavity photon leaks out as
$\ket{\Psi_{\mu-1},1,0}\rightarrow\ket{\Psi_{\mu-1},0,0}$. This energy-lowering
process occurs successively until the population reaches the ground state. 
This cooling mechanism applies for any number of atoms $N$.

Note that this simple model already illustrates a novel application of the
present mechanism: the ground state, which is a maximally entangled state, is
prepared as a steady state without any initialization of the system or time
control of the fields. For a small number of atoms, one can apply several
resonant fields in the same manner to prepare the entangled ground state. The
effective spontaneous emission rate $\sim\sum_j\gamma|\Omega_j/\Delta|^2$,
with $\gamma$ the intrinsic spontaneous emission rate, can be made
arbitrarily negligible compared to the Raman transition frequency
$\sim\sum_j|g_j\Omega_j/\Delta|$ {by increasing $\Delta$}.

\section{General scheme}

The above analysis leads us to the reasoning that
if the classical fields are broadband with $\delta_{1,2}>0$, many-body
systems would tend to cool even if they have many energy levels with
different energy gaps, hence many different resonance conditions. Let us
first simplify the situation by considering spectrally incoherent broadband
fields. We denote the spectral density of the broadband field by $I(\delta)$,
which is normalized as $\int{I(\delta)}d\delta=1$ (the subscripts 1
and 2 {are omitted} for brevity). The spectral incoherence can be reflected by replacing
$\Omega_je^{i\delta{t}}$ with $\Omega_j\int{d\delta}f(\delta){e}^{i\delta{t}}$ and
$\abs{\Omega_j}^2\int{d\delta}d\delta'f(\delta)f^*(\delta')$ explicitly with
$\abs{\Omega_j}^2\int{d\delta}d\delta'I(\delta)\mathfrak{d}(\delta-\delta')$, where
$\mathfrak{d}(x)$ is the Dirac delta function. The standard treatment of the
time-dependent perturbation theory then yields the Fermi-golden-rule-like
transition rate $2\pi{n_1}\abs{(\Gamma_-)_{\mu\nu}}^2I(E_{\nu\mu})$ for
transition $\ket{\Psi_\nu,n_1-1,n_2}\rightarrow\ket{\Psi_\mu,n_1,n_2}$
(similarly for $n_2$).
$H_d$ is invariant under the above change.

Now that the transition is understood to be Markovian, it can be represented by a Markov chain, where each node represents an
energy level (including the cavity state). For each pair of the energy
levels, the transition is established if the broadband field contains the
spectral component $\delta$ corresponding to the energy gap. The cavity decay
$\ket{\Psi_\mu,n_1,n_2}\rightarrow\ket{\Psi_\mu,n_1-1,n_2}$ is also included
with rate $n_1\kappa$ (similarly for $n_2$).

The detailed relaxation process depends on the specific many-body systems at
hand. In what follows, we discuss the relaxation process of the isotropic
Heisenberg spin chain with Hamiltonian
$H_0=J\sum_{j=1}^{N-1}\vec{s}_{j}\cdot\vec{s}_{j+1}$, where $J>0$ and $N$ is
even for $\ket{\Psi_0}$ to be nondegenerate. Those arguments below, however,
mostly hold when $H_0$ commutes with the total spin operator
$S_z=\sum_j{s_j^z}$, as is the case for any Heisenberg spin system without XY
anisotropy. Most of the spin systems simulable by earlier schemes fall into
this class. Several remarks are in order. (i) All the conditions mentioned
above are inherited. However, we strictly impose them only between
$\ket{\Psi_0}$ and every $\ket{\Psi_\mu}$ with $\mu\neq0$, since this is
important for the ground state to remain stable. Even if the conditions are
not met for some higher levels, the overall effect still pushes the
population downwards and once it reaches the ground state, any further
transition is suppressed. (ii) We lift most degeneracies by
applying a magnetic field $BS_z$, which can be done by an
additional field along with the $\epsilon_z$ term. As $[H_0,S_z]$, the
eigenstates of $H_0$ are now the simultaneous eigenstates of $S_z$.
$\ket{\Psi_0}$ remains unchanged. (iii) We break the spatial coherence by
applying spatially incoherent fields, or alternatively by breaking the
symmetry, e.g., by tiling, focusing, or displacing the fields and applying
those complementary field configurations in turn. In what follows, we simply
assume
$|(\Gamma_\pm)_{\mu\nu}|^2=\sum_j|\frac{g_j\Omega_j}{\Delta}(s_j^\pm)_{\mu\nu}|^2$.

\begin{figure}
	\includegraphics[width=0.9\columnwidth]{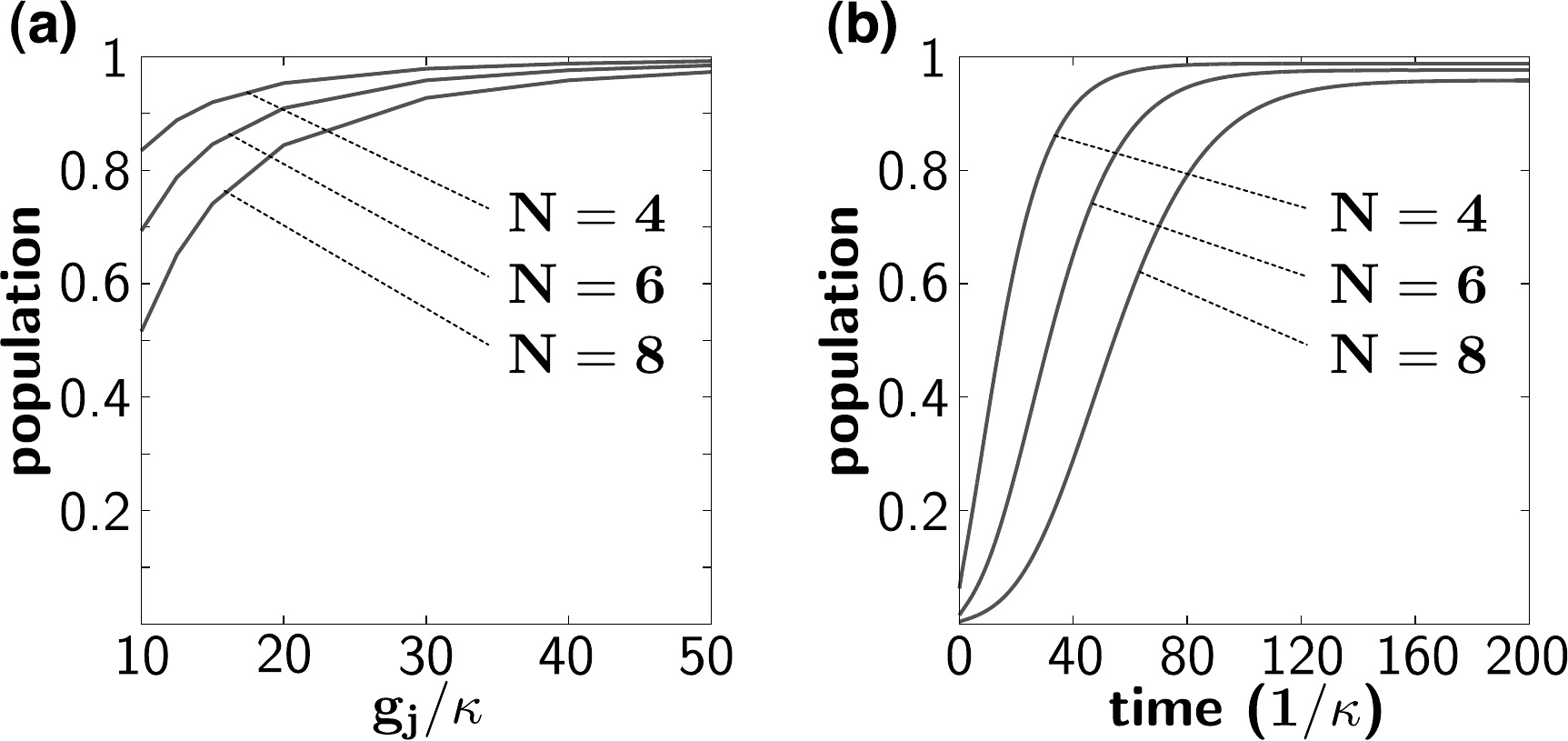}
	\caption{Numerical simulation for Heisenberg spin chains. (a) The asymptotic population in the ground state. (b) The time evolution of the ground state population for $g_j=40\kappa$.}
	\label{fig:general}
\end{figure}

\section{Efficiency}

The dependency of the relaxation time on $N$ is an
important issue. If it grows exponentially with $N$, the relaxation scheme
would be less useful. Although it is daunting to obtain a general
relationship, one can see that it grows at most polynomially with $N$ for
gapped systems as explained below. 
Note that this is related to the fundamental
limit originated from the energy-time uncertainty.
If there exists no energy gap between the ground state and the excited states, it is not 
possible to resolve the ground state within a finite time scale, hence the ground state
cooling is possible only approximately at best. Fortunately, the ground states of thermodynamically gapped systems---Haldane gap systems, quantum Hall systems, superconductors, etc.---are usually of special interest, for which our underlying idea may be applicable in relevant forms. The Heisenberg chain we analyze here is thermodynamically gapless.

The relaxation time can be estimated as the number of
energy levels to pass through (in the Markov chain) to reach the ground
level, say $N_S$, multiplied by the characteristic transition time
$\sim1/2\pi|(\Gamma_\pm)_{\mu\nu}|^2I(\delta)$ between two levels. Although
the coefficient $|(\Gamma_\pm)_{\mu\nu}|^2$ is bounded by $E_{10}$, it can be
kept finite for gapped systems. Besides this, we end up with a trade-off
problem: as the fields get broader-band, $N_S$ decreases while $1/I(\delta)$
increases. By noticing that the eigenstates of $H_0$ are also the eigenstates
of $S_z$ (ranging from $-N/2$ to $N/2$) and a transition changes $S_z$ by
one, we see that if the broadband covers the entire energy spectrum, $N_S$ is
at most $2N$ ($N$ for spin flips and $N$ for cavity decays), while
$1/I(\delta)$ behaves as $N$.

\section{Spontaneous emission}

Although the atomic excitation is highly
suppressed, the spontaneous emission is the prominent source of heating. The
spontaneous emission changes the state as
$\ket\Psi\rightarrow{s_j^\pm}\ket\Psi$ or ${s_j^\pm}s_j^\mp\ket\Psi$, which
can be reflected in the Markov chain by adding transitions
$\ket{\Psi_\mu}\leftrightarrow\ket{\Psi_\nu}$ with rates
$\sum_j(\gamma/4)|\Omega_j/\Delta|^2\bigl[|(s_j^+)_{\mu\nu}|^2+|(s_j^-)_{\mu\nu}|^2+2|(s_j^z)_{\mu\nu}|^2\bigr]$,
where the branching ratio is assumed to be $50:50$. 
{Note that when spectrally incoherent broadband fields are used, the transition rate is of the same order
$\sim|\Omega_j/\Delta|^2$ as the spontaneous emission. In such a case, strong atom-cavity coupling $g_j\gg\gamma$ is required to obtain a high fidelity to the ground state. See the numerical simulation section below.} 
On the other hand, when spectrally coherent broadband fields, i.e., pulses, are used,
the spontaneous emission again can be made negligible, in principle, 
by increasing $\Delta$, as in the previous case of Fig.~\ref{fig:toy}. 
In this case, only moderately strong atom-cavity coupling should suffice.
If the spin
system is realized with trapped atoms, the spontaneous emission may also
excite the atomic momentum. This could be overcome by sympathetic
cooling~\cite{larson:1986a, daley:2004a, griessner:2004a}.

\section{Numerical simulation}

In Fig.~\ref{fig:general}a, we plot the asymptotic ground state population with respect to $g_j$ for $N=\{4,6,8\}$.
We take $\Delta/g_j=g_j/\kappa$ and $\kappa=\gamma=B/10$ with $B=E_{10}/2$ (note that $E_{10}$ depends on $J$ and $N$). 
$\Omega_j$ is chosen so that $|(\Gamma_+)_{10}|=\kappa$. $I(\delta)$ is constant for
$0.5B<\delta-(\epsilon_1)_{00}<3.5B$ and zero otherwise. 
In Fig.~\ref{fig:general}b, we plot the time evolution of the population in the ground state starting from the maximally mixed state for $g_j=40\kappa$. 
Note that the spectral incoherence, introduced just for ease of analysis and numerical simulations, makes the process inefficient, for which a rather strong atom-cavity coupling is required. As explained above, if  spectrally coherent fields are used, this demanding requirement should be mitigated. 

\section{Experimental regimes}

As the typical energy scales $(J,B)$ of current quantum simulation experiments and proposals range from 1 to 100~kHz~\cite{duan:2003a,garcia-ripoll:2004a,sim11,porras:2004a,friedenauer:2008a,kim10,bri12,isl13,hartmann:2007a,cho:2008a}, the most crucial requirement is the condition $\kappa\ll E_{10}$. A 1-10~kHz decay rate for an optical cavity is experimentally accessible \cite{wol12}. Once this condition is met, others can be satisfied by adjusting field intensities and detunings. As cavity QED parameters scale roughly as $\kappa\propto1/L$ and $g\propto1/\sqrt{L}$ with $L$ the cavity length, $\kappa$ can be decreased without sacrificing the ratio $g/\kappa$ \cite{hood:2001a}. One may also exploit, e.g., the Feshbach resonance~\cite{inouye:1998a}, by which the characteristic energy scale of the quantum simulator itself can be drastically increased.

\section*{Acknowledgement}

This work was supported by the UK Engineering and Physical Sciences Research Council, Royal Society, Wolfson Foundation, and National Research Foundation \& Ministry of Education Singapore.


\end{document}